\documentclass[12pt]{article}
\usepackage{graphicx}
\usepackage{epsfig}
\begin{document}
\title{A simple adaptive-feedback controller for identical chaos synchronization}
\author{{
Debin Huang \footnote{E-mail: dbhuang@staff.shu.edu.cn}
}\\\\
{\small Department of Mathematics, Shanghai University, Shanghai
200436, P.R. China}}
\date{}
\maketitle
\begin{center}
\begin{minipage}{12cm}
\hskip 1cm  Based on the invariance principle of differential
equations, a simple adaptive-feedback scheme is proposed to
strictly synchronize almost all chaotic systems. Unlike the usual
linear feedback, the variable feedback strength is automatically
adapted to completely synchronize two almost arbitrary identical
chaotic systems, so this scheme is analytical, and simple to
implement in practice. Moreover it is quite robust against the
effect of noise. The famous Lorenz and R$\ddot{\textrm{o}}$ssler
hyperchaos systems are used as
illustrative examples.\\
\vskip 0.5cm PACS number(s): 05.45.Xt, 87.17.Nn
\end{minipage}
\end{center}
\hskip 0.5cm Since it  was shown in [1] that for some chaotic
systems the synchronization is possible, synchronization of
(unidirectionally) coupled chaotic systems and its potential
applications in engineering have been a field of great interest
over a decade, see [2-4] and references cited therein. Due to the
different applications, various specific synchronization schemes
have been proposed in the literature, see [3] and references cited
therein. However, just as what stated in [5], despite the large
amount of effort many key issues remain open. One of the central
questions is: Given two arbitrary identical chaotic systems, how
can one design a physically available coupling scheme that is
strictly guaranteed to produce stable identical synchronization
motion (i.e., high-quality synchronization)? In most of the
rigorous results based on the Lyapunov stability or the linear
stability, the proposed scheme is very specific, but also the
added controller is sometimes too big to be physically practical.
One practical scheme is the linear feedback. However, in such
technique it is very difficult to find the suitable feedback
constant, and thus numerical calculation has to be used, e.g., the
calculation of the conditional Lyapunov exponents. Due to
numerical calculation, such scheme is not regular since it can be
applied only to particular models. More unfortunately, it has been
reported that the negativity
 of the conditional
Lyapunov exponents is not a sufficient
 condition for complete chaotic synchronization,
 see [6]. Therefore the synchronization based on these numerical schemes
 can not be strict (i.e., high-qualitative), and is generally
  not robust against the effect of noise. Especially, in these schemes a very weak noise or a small parameter mismatch
  can trigger the desynchronization problem due to a sequence of bifurcations [7].\par
 Actually, this open problem, although significant for complete chaos synchronization, is very difficult and cannot admit the
 optimization solution [3]. For example, in [5] a rigorous criteria is presented
 to guarantee linearly stable synchronization motion, but the criteria is
  so complicated that the specific numerical calculation is
  necessary for particular examples in practice. The similar problem was ever addressed in
 [8].
\par
 In this letter, we give a novel answer to the above open problem. We prove rigorously
 by using the invariance principle of differential equations [9] that a simple
 feedback coupling
  with the updated feedback strength, i.e., an adaptive-feedback scheme, can strictly synchronize
 two almost arbitrary identical chaotic systems.
 \par Let a chaotic (drive) system be given as
$$\dot x=f(x), \eqno(1)$$
where $x=(x_1,x_2,\cdots,x_n)\in \textrm{R}^n$, $f(x)=(f_1(x),
f_2(x),\cdots,f_n(x)):\textrm{R}^n\rightarrow \textrm{R}^n$ is a
nonlinear vector function. And let $\Omega\subset\textrm{R}^n$ be
a chaotic bounded set of (1) which is globally attractive. For the
vector function $f(x)$, we give a general assumption. \par {\sl
For any $x=(x_1,x_2,\cdots, x_n)$, $x_0=(x_1^0,x_2^0,\cdots,
x_n^0)\in \Omega$,
 there exists a
constant $l>0$ satisfying $$\mid f_i(x)-f_i(x_0)\mid\leq l
\textrm{max}_j\mid x_j-x_j^0\mid, i=1,2,\cdots, n.\eqno (2)$$}
\par
We call the above condition as the uniform Lipschitz condition,
and $l$ refers to the uniform Lipschitz constant. Note this
condition is very loose, for example, the condition (2) holds as
long as ${\partial f_i\over
\partial x_j} (i,j=1,2,\cdots, n)$ are bounded. Therefore the class of systems in the form of (1)-(2)
include all well-known chaotic and hyperchaotic systems. Consider
the variables of (1) as coupling signals, the receiver system with
variables $y\in \textrm{R}^n$ is given by the following equations:
$$
\dot y=f(y)+\epsilon (y-x), \eqno(3)
$$
where the feedback coupling $ \epsilon
(y-x)=(\epsilon_1e_1,\epsilon_2e_2,\cdots, \epsilon_ne_n)$,
$e_i=(y_i-x_i),i=1,2,\cdots,n$ denoting the synchronization error
of (1) and (3). Instead of the usual linear feedback, the feedback
strength $\epsilon=(\epsilon_1,\epsilon_2,\cdots, \epsilon_n)$
here will be duly adapted according to the following update law:
$$
\dot {\epsilon}_i=-\gamma_ie_i^2,i=1,2,\cdots,n,\eqno(4)
$$
where $\gamma_i>0, i=1,2,\cdots,n,$ are arbitrary constants. For
the system of $2n$ equations (which is formally called as the
augment system for the convenience below), consisting of the error
equation between (1) and (3), and the equation (4), we introduce
the following nonnegative function
$$
V={1\over 2}\sum_{i=1}^ne_i^2+{1\over
2}\sum_{i=1}^n{1\over\gamma_i}(\epsilon_i+L)^2, \eqno(5)
$$
where $L$ is a constant bigger than $nl$, i.e., $L>nl$. By
differentiating the function $V$ along the trajectories of the
augment system, we obtain
$$
\dot V=\sum_{i=1}^ne_i(\dot {y}_i-\dot
{x}_i)+\sum_{i=1}^n{1\over\gamma_i}(\epsilon_i+L)\dot \epsilon$$
$$
=\sum_{i=1}^ne_i(f_i(y)-f_i(x)+\epsilon_ie_i)-\sum_{i=1}^n(\epsilon_i+L)e_i^2
\leq(nl-L)\sum_{i=1}^ne_i^2\leq 0. \eqno (6)
$$
where we have assumed $x,y\in \Omega$ (without loss of the
generality as $\Omega$ is globally attractive ), and used the
uniform Lipschitz condition (2). It is obvious that $\dot V=0$ if
and only if $e_i=0, i=1,2,\cdots, n$, namely the set
$E=\{(e,\epsilon)\in \textrm{R}^{2n}:e=0,\epsilon=\epsilon_0\in
\textrm{R}^{n}\}$ is the largest invariant set contained in $\dot
V=0$ for the augment system. Then according to the well-known
invariance principle of differential equations [9], starting with
arbitrary initial values of the augment system, the orbit
converges asymptotically to the set $E$, i.e., $y\rightarrow x$
and $\epsilon\rightarrow \epsilon_0$ as $t\rightarrow\infty$.
\par
Obviously, such identical synchronization motion is strict (i.e.,
high-qualitative), global (as long as the chaotic attractor is
globally attractive), and nonlinear stable. In particular the
nonlinear global stability implies that such chaos synchronization
is quite robust against the effect of
 noise, namely under the case of presenting a small noise the synchronization
 error eventually approaches zero and ultimately fluctuates around zero wheresoever the initial values start. In addition, we note that in order to reach the synchronization
the variable feedback strength
 $\epsilon$ will be automatically adapted to a suitable strength $\epsilon_0$ depending on the initial
 values, which is significantly different from the usual linear feedback.
 As is well known, in the usual linear feedback scheme
 a fixed strength is used wheresoever the initial values
 start, thus the strength must be maximal, which means a kind of
 waste in practice. However the final strength in the present scheme
 depends on the initial error, thus the strength is generally equal or less than those used in the constant gain schemes.
 Moreover, the present control scheme does not require to
 determine numerically any additive parameters, and is simple to
 implement in practice since the technique is similar to the
 well-known self-adaptive controller in the control theory.
 Note that although the converged strength is not bigger than the linear one, theoretically
 the strength may be too big to be practical. However the
 flexibility of the strength in the present scheme can overcome
 this limitation once such case arises. For example, suppose that the
 feedback strength is restricted not to exceed a critical value,
 say $k$. In the present control procedure, once the variable strength $\epsilon$ exceeds $k$ at
 time $t=t_0$, we may choose the values of variables at this time
 as initial values and repeat the same control by resetting the
 initial strength $\epsilon(0)=0$. Namely one may achieve the
 synchronization within the restricted feedback strength due to
 the global stability of the present scheme, which is slightly similar to
 idea of OGY control [10]. This excellence
 is absent from the usual feedback scheme. In addition, in the present scheme the small converged
 strength may be obtained by decreasing suitably the value of parameter
 $\gamma$, which governs the rate of increasing of the feedback strength. As for the question on how the parameter $\gamma$
affects the convergence rate (i.e., transient time) and the final
coupling strength remains to be further investigated.
 \par
Note that in the proposed scheme the coupling all the variables
may be redundant to achieve synchronization, because we find from
my proof that it is not necessary, e.g., one may set
$\epsilon_i\equiv0$ (i.e., cancelling the corresponding coupling)
if $\mid e_i\mid\leq\mid e_j\mid$. Actually this case exists in
general due to the nonhyperbolicity of chaotic attractor, namely
near a nonhyperbolic point the divergence rate of trajectories
(resp. the contraction and stretching of phase space along
trajectories) in some directions vanishes (or is very small), see
the following examples. Although we can not give a rigorous
criteria to determine only which variables can be used as the
coupling signals for the general examples, in a concrete model one
may determine it by the numerical technique, e.g., the calculation
of the Lyapunov exponent respective to each direction. Of course
for the low dimensional systems the optimal coupling variables may
be found by directly testing again and again. Especially, if based
on the calculation of the conditional Lyapunov exponents a chaotic
system can be synchronized by linearly coupling a variable, then
the coupling this variable in the present scheme can surely
achieve synchronization, see the first example below.

\par
 Next we will give two illustrative examples. Consider the Lorenz
 system,
 $$\dot{x}_1=\beta(x_2-x_1), \ \ \ \dot{x}_2=\alpha x_1-x_1x_3-x_2,\ \ \
 \dot{x}_3=x_1x_2-b x_3.
 \eqno(7)$$
 The corresponding receiver system is:
$$
\dot{y}_1=\beta(y_2-y_1)+\epsilon_1 e_1, \ \ \dot{y}_2=\alpha
y_1-y_1y_3-y_{2}+\epsilon_2e_2,\ \
\dot{y}_3=y_{1}y_{2}-by_{3}+\epsilon_3e_3 \eqno(8)$$
 with the update law (4). Now let $\beta=10,
\alpha=28,b={8\over 3}$. It has been well known that
 a suitable linear feedback in the second component may synchronize the Lorenz system, so
 we let $\epsilon_1=\epsilon_3\equiv 0$ (i.e., the time series of only the
variable $x_2$ are selected as the driven signal), and set
$\gamma_2=0.1$. The corresponding numerical results are shown in
Figure 1 and Figure 2, where the initial feedback strength is set
as zero. Figure 1 shows the temporal evolution of synchronization
error between (7) and (8) , and the variable strength
$\epsilon_2$. Figure 2 shows that when an additive uniformly
distributed random noise in the rang $[-1,1]$ (i.e., a noise of
the strength $1$) is present in the signal output $x_2$ of (7),
the synchronization error eventually approaches zero and
ultimately fluctuates slightly around zero, meanwhile the variable
feedback strength is affected slightly and can not stabilize.
\par As the second example, we consider the
R$\ddot{\textrm o}$ssler hyperchaos system:
$$
  \dot {x}_1=-x_2-x_3,\ \ \ \ \dot {x}_2=x_1+0.25x_2+x_4,$$
 $$\dot {x}_3=3+x_1x_3,\ \ \ \  \dot {x}_4=-0.5x_3+0.05x_4.
 \eqno(9)$$
 The receiver system is:
$$
\dot{y}_1=-y_2-y_3+\epsilon_1e_1, \ \ \ \
\dot{y}_2=y_1+0.25y_2+y_4+\epsilon_2e_2,$$
$$\dot{y}_3=3+y_1y_3+\epsilon_3e_3,\ \ \ \ \dot {y}_4=-0.5y_3+0.05y_4+\epsilon_4e_4 \eqno(10)$$
with the update law (4).
 Let $\epsilon_1\equiv0$ (we speculate
 that this is not optimal),
$\gamma_i=1,i=2,3,4$, and initial feedback strength  be $(0,0,0)$,
numerical results for two different cases are shown in Figure 3
and Figure 4, respectively. Figure 3 shows the hyperchaotic
synchronization and Figure 4 the slight effect of a noise with the
strength $0.1$ which is simultaneously added to the signals $x_2$,
$x_3$ and $x_4$.
\par

The above numerical examples show that chaotic or hyperchaotic
synchronization can be quickly achieved by the present controller
(i.e., the transient time to synchronization is very short). In
addition, we find from these examples that such synchronization is
robust against the effect of noise, namely if the expected
synchronization means that the synchronization error is eventually
smaller than a threshold value (not tends to zero), then the
present scheme is physically feasible in the noisy case. Moreover,
by comparing the converged feedback strength and the corresponding
feedback signals we find that the coupling is indeed small in the
examples. In addition by testing the other chaotic systems
including the R$\ddot{\textrm o}$ssler system, Chua's circuit, and
the Sprott's collection of the simplest chaotic flows we find that
the coupling only one variable is sufficient to achieve identical
synchronization of a three-dimensional system.
\par In conclusion, we have given a novel answer to the open problem in the field of identical chaos
synchronization. In comparison with the previous methods, the
proposed scheme supplies a simple, analytical, and (systematic)
uniform controller to synchronize strictly two arbitrary identical
chaotic systems satisfying a very loose condition. The technique
is simple to implement in practice, and quite robust against the
effect of noise. The control idea may be also generalized to the
case of the discrete chaotic systems. We also believe that such
simple synchronization controller will be very beneficial for the
applications of chaos synchronization. Especially the similar
control scheme has been successfully used to stabilize the chaotic
neuron model [11], and hence the proposed adaptive-feedback
synchronization controller can be used to explore more reasonably
the interesting dynamics found in neurobiological systems, i.e.,
the onset of regular bursts in a group of irregularly bursting
neurons with different individual properties [12].
\par
 {\bf Acknowledgments:} The author is in debt to all referees
  including those consulted by PRL and PRE for their valuable suggestions.
  This work is supported by the National Natural Science
Foundation of China (10201020; 10432010). \small {

}

\begin{figure}
\begin{center}\epsfig{file=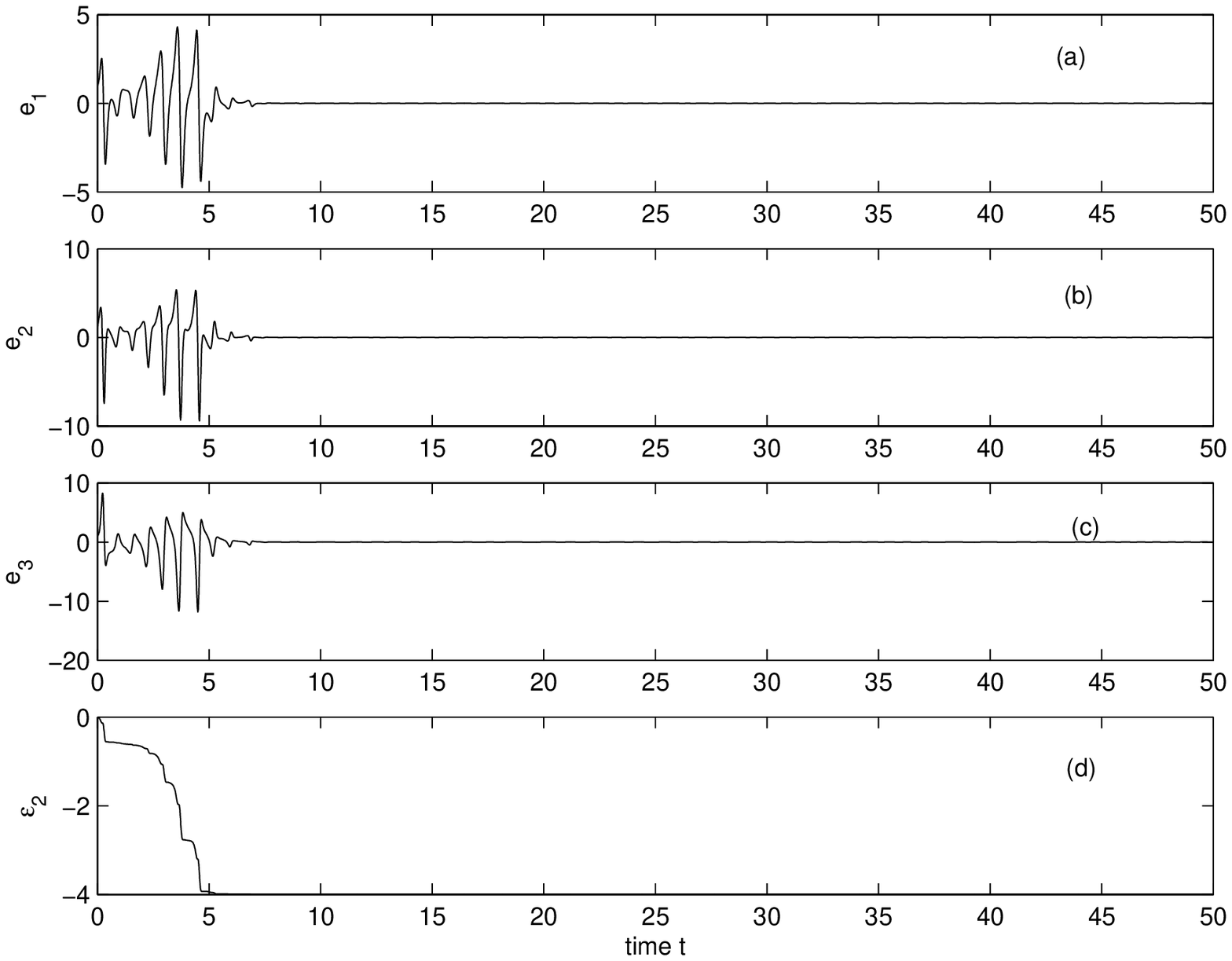,height=8cm,width=8cm}\end{center}
 \begin{center}\begin{minipage}{8cm}{\footnotesize {\bf FIG.1}.
The chaos synchronization between (7) and (8) is achieved by only
the signal $x_2$,
 where (a)-(c) show temporal evolution of the synchronization error
 and (d) shows the evolution of the corresponding feedback strength $\epsilon_2$. Here the initial values of $(x,y,\epsilon)$ is set as $(2,3,7,3,4,8,0)$. }\end{minipage}\end{center}
\end{figure}

\begin{figure}
\begin{center}\epsfig{file=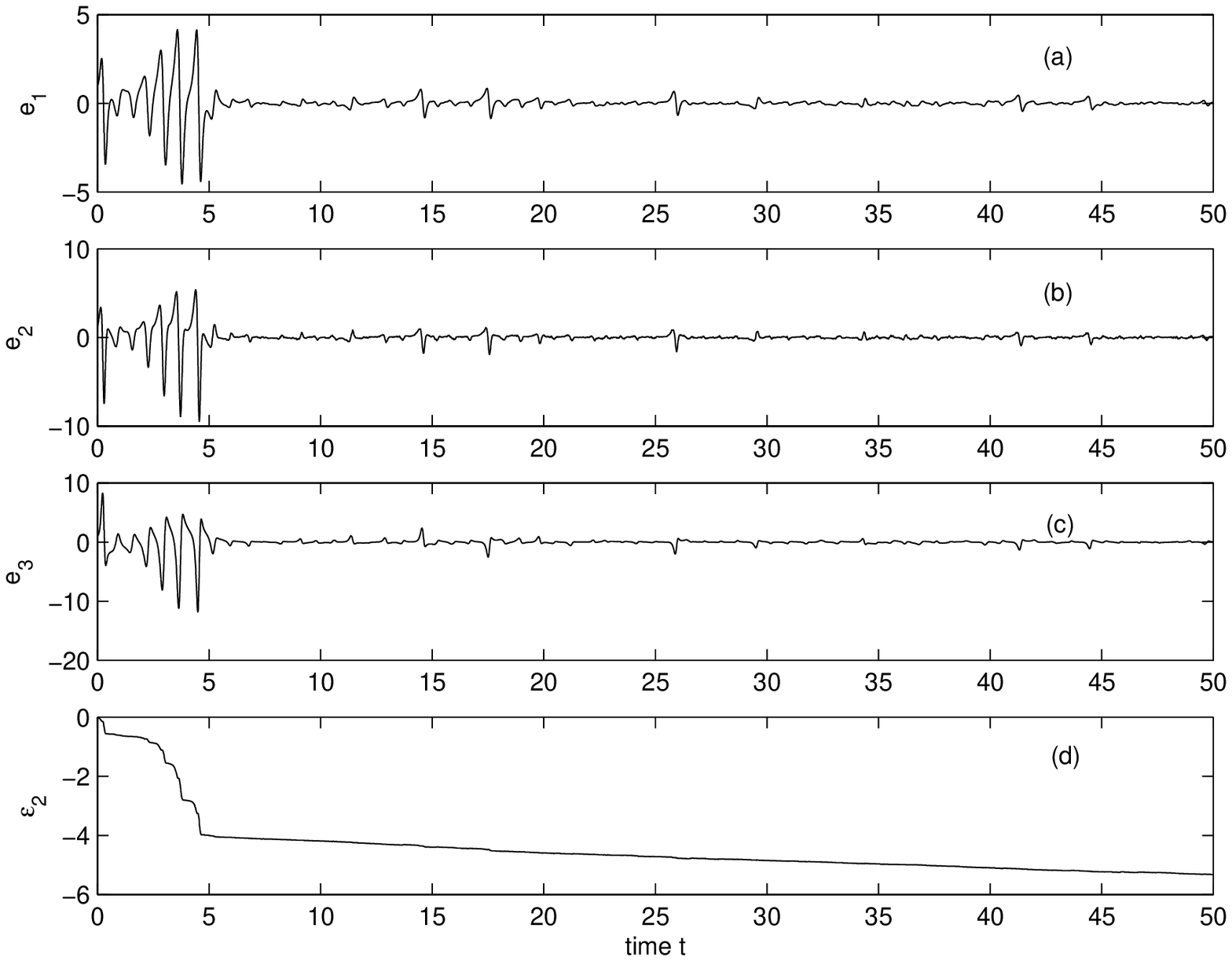,height=8cm,width=8cm}\end{center}
 \begin{center}\begin{minipage}{8cm}{\footnotesize {\bf FIG.2}.
(a)-(d) show the effect on  the synchronization error and the
feedback strength in Figure 1 when a noise
 of
the strength $1$ is present in the signal output $x_2$ of
(7).}\end{minipage}\end{center}
\end{figure}

\begin{figure}
\begin{center}\epsfig{file=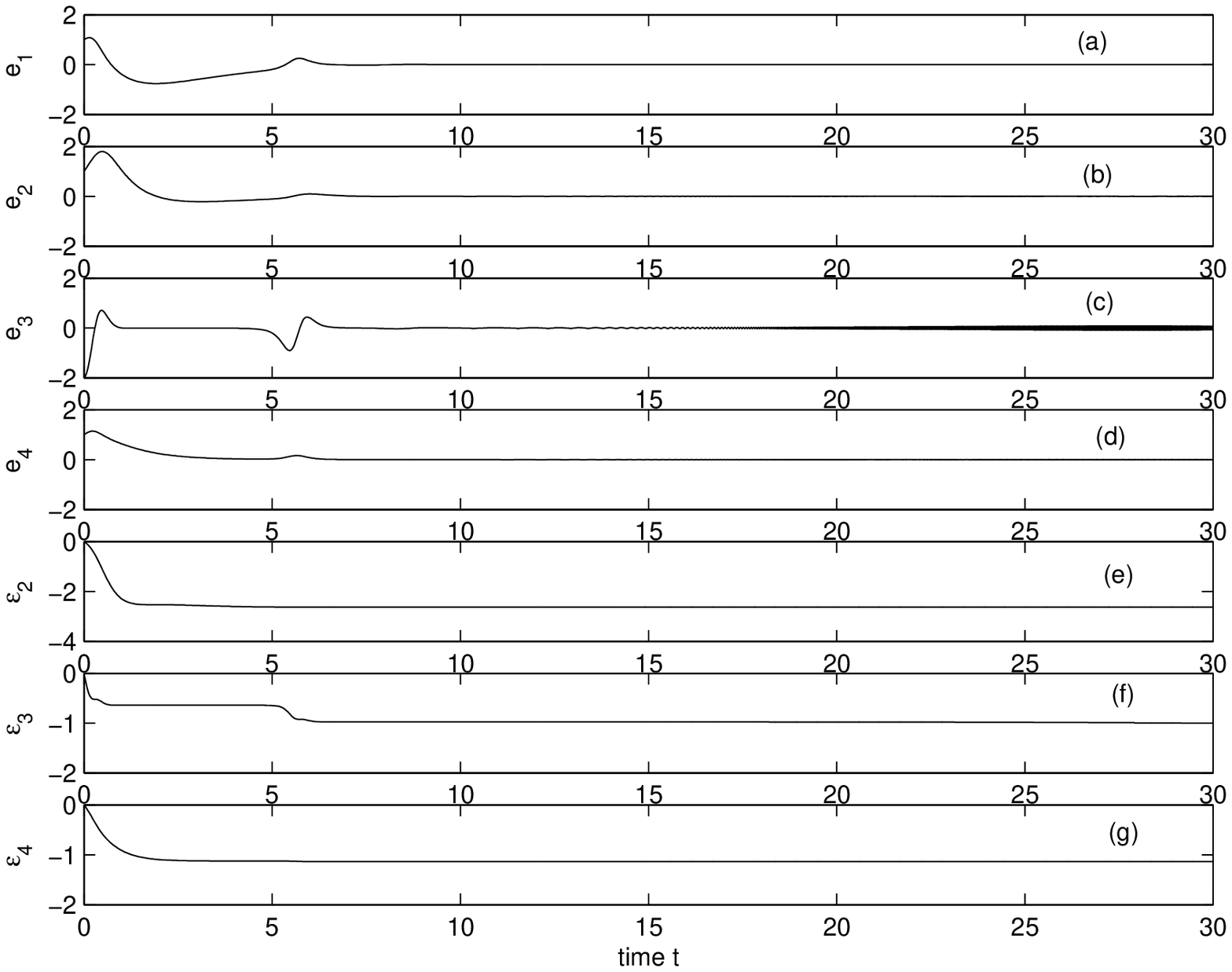,height=8cm,width=8cm}\end{center}
 \begin{center}\begin{minipage}{8cm}{\footnotesize {\bf FIG.3}.
(a)-(g) show the hyperchaotic synchronization between  systems (9)
 and (10), and temporal evolution of the corresponding feedback strength,
 where only three variables of (9), $x_i,i=2,3,4$ are selected as the driven signals. Here the initial values of $(x,y,\epsilon)$ is set as $(2,3,7,10,3,4,5,11,0,0,0)$}\end{minipage}\end{center}
\end{figure}

\begin{figure}
\begin{center}\epsfig{file=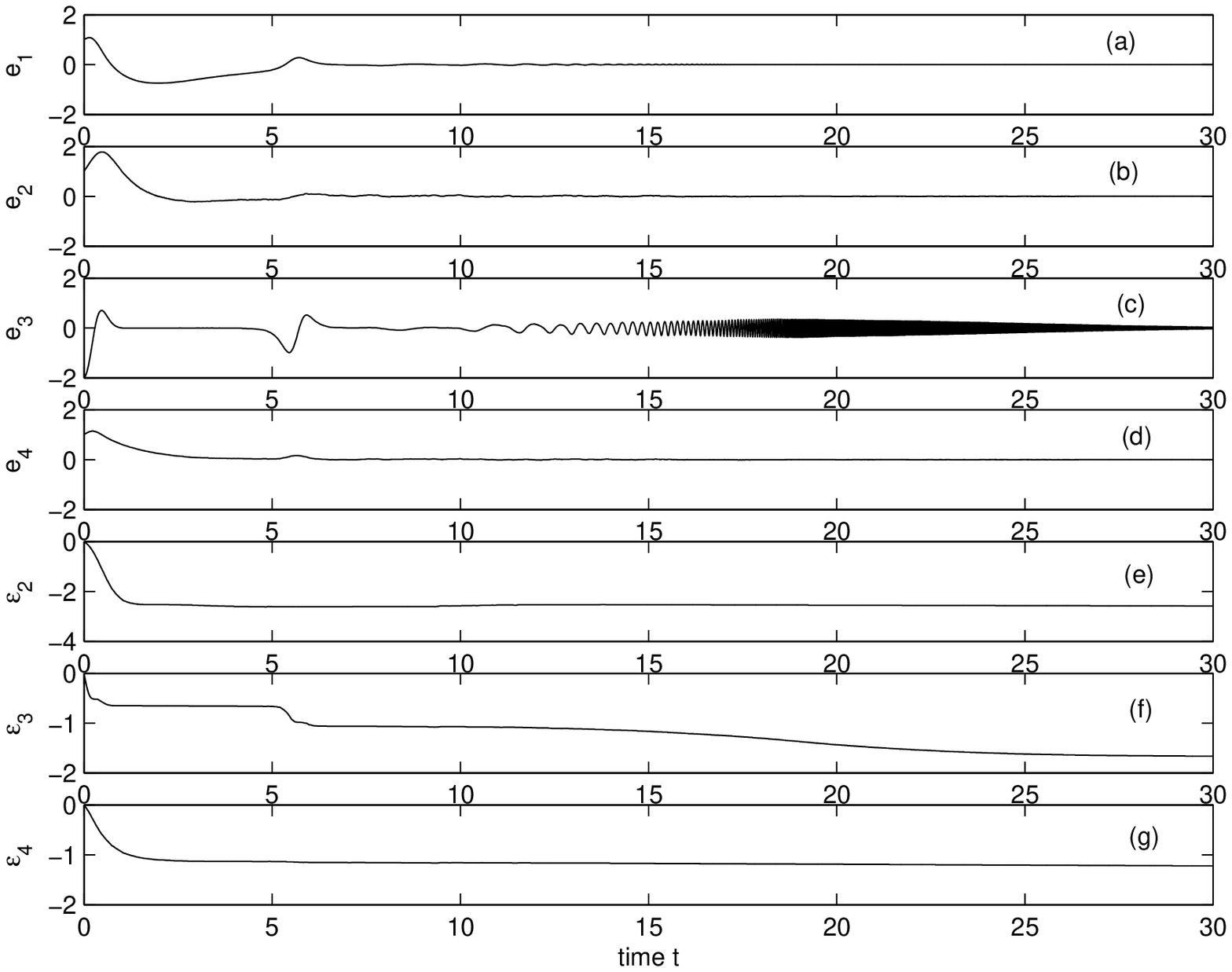,height=8cm,width=8cm}\end{center}
 \begin{center}\begin{minipage}{8cm}{\footnotesize {\bf FIG.4}.
(a)-(g) show the effect on  the synchronization error and the
feedback strength in Figure 3 when a noise
 of
the strength $0.1$ is simultaneously added to the signal outputs
$x_2$, $x_3$ and $x_4$ of (9).}\end{minipage}\end{center}
\end{figure}

\end{document}